\documentclass[12pt]{article}
\usepackage{epsfig}
\usepackage{array}
\usepackage{amsmath}
\usepackage{amssymb}

\newcommand{\be}{\begin{equation}}
\newcommand{\ee}{\end{equation}}
\newcommand{\bea}{\begin{eqnarray}}
\newcommand{\eea}{\end{eqnarray}}

\begin{document}

\title{ Shake-off in the $^{164}$Er neutrinoless double
electronic capture and the dark matter puzzle }
\author{ F. F. Karpeshin\footnote{e-mail: fkarpeshin@gmail.com} \\ Mendeleev All-Russian Research
Institute of Metrology \\ 190005 Saint-Petersburg, Russia \\
\bigskip
and \\ M. B. Trzhaskovskaya\footnote{Deceased} \\
Petersburg Nuclear Physics Institute 
of the National Research Center \\ ``Kurchatov Institute'' \\
188300 Gatchina, Russia }

\maketitle

\begin{abstract}
Traditionally double neutrinoless electronic capture is considered as a
resonance process. We have fulfilled shake-off probability
calculations, leading to ionization of the electron shell, in the
case of $^{164}$Er. Allowance for the shake-off removes the requirement
of resonance, leading to a radical increase of the capture rate. In the case of $^{164}$Er, the
contribution of the new mechanism increases the capture rate by a
factor of 5.6 as compared to the conventional resonance
fluorescence mechanism. It also increases the probability of
electron capture from higher shells, which must be foreseen in the
experimental studies. Moreover, effect of the shake-off can
expand the list of candidate nuclei for experimental research. 
The influence of the shake-off is also expected to manifest itself in the other $\beta$ processes which are used in the studies of the neutrino nature, its mass and role in the dark matter puzzle. 
\end{abstract}
\maketitle

\large
\bigskip

      \section{Introduction}
      The discovery of dark matter in the universe stimulates the development of theories beyond the Standard Model. As a rule, they include violation of the lepton quantum number, unless special restrictions are introduced. This attracts great interest in the study of double beta processes, including the $2e$ decay of a nucleus and the capture of two orbital electrons by it \cite{obzor}. Within the framework of the Standard Model, the lepton quantum number is conserved. This excludes double neutrinoless beta decay or $e$ capture. The latter become possible only if neutrinos have mass and
if neutrinos are particles of the Majorana nature. However, the
discovery of mass in neutrinos and their oscillations has already
marked the observation of processes beyond the Standard Model.
Thus, the search for neutrinoless binary processes should answer
the question about the Majorana nature of neutrinos. Of the two
neutrinoless processes, the $2e0\nu$ decay has the highest decay
rate. The $2e0\nu$ capture, although inferior to the $2e0\nu$
decay by several orders of magnitude in probability, is more
convenient from the viewpoint  of detection.

      The essential point is that neutrinoless $2e$ capture has traditionally been considered as a resonance process, since not a single particle is emitted as a result of nuclear transformation \cite{Wyceh}. At the same time, the conservation law  requires the transfer of the part of the released energy to a third body. This is the electron shell of the atom. The law of energy conservation  is restored, for example, due to the emission of a fluorescence quantum, the energy of which includes the excessive $Q$ value. Another reason for the shift is formation of the two vacancies in the places of the captured electrons and thus inflated in comparison with a normal atom electron shell. Therefore, the energy of these photons differs from ordinary fluorescence photons, as usually fluorescence  takes place in ionized atoms with a single hole in the inner shell. The detection of such satellites in the fluorescence spectrum can serve as an indicator of either neutrinoless or two-neutrino double electron capture \cite{h2,FK}. Thus, the neutrinoless capture amplitude permanently includes the radiative vertex, which retards the process by two orders of magnitude. Therefore, the main criterion is focused on the study of nuclei with the small $Q$ values. In Ref. \cite{elis}, a list of three most suitable candidates for experimental research was selected: $^{152}$Gd, $^{164}$Er, and $^{180}$W. In this paper, we refine the question of the probability of the $^{164}$Er $\to ^{164}$Dy process based on the new shake-up mechanism, which was  proposed in \cite{nrl}.

      The shake-off  mechanism does not require a resonance. For this reason, its contribution decreases more slowly with increasing the resonance defect $\Delta$ than the conventional resonant fluorescence mechanism. The restoration of the energy conservation law  occurs due to the ionization of the electron shell. In this case, the excess energy is carried away by the ejected electron. Estimates of the shake-off efficiency were made in \cite{nrl} using the example of the $^{152}$Gd $\to$ $^{152}$Sm decay, which has the smallest resonance defect of $\Delta$ = 0.910 keV among the known candidates. Quantitatively, the contribution of the new mechanism turned out to be at the level of 23\% as compared to the traditional mechanism. In this paper, we consider the $^{164}$Er $\to$ $^{164}$Dy process,  with a higher value of $\Delta$ = 6.82 keV. The results confirm expectations. Accounting for shake-off reduces the expected half-life of the process by almost 6 times. In the next section, we recall the basic formulas. The calculation results are given in Sec. 3. Section 4 is devoted to a discussion of the results obtained in this work.

      \section{Comparison of the two mechanisms of neutrinoless $2e$ capture: physical principles and formulae for calculation}

      In the case of $2e0\nu$ capture, the atom remains generally neutral. Therefore, the energy release is determined by the difference in the masses of neutral atoms,  the initial $M_1$ and the daughter one $M_2$ (we use the relativistic system of units $\hbar = c = m_e = 1$, with $m_e$ being the electron mass, unless otherwise noted):
      \be
      Q=M_1-M_2\,.  \label{Q}
      \ee

      However, the process with the total energy release (\ref{Q}) could be realized only with the capture of the outermost, valence, electrons. As a rule, the capture of internal electrons is much more probable, as their density on the nucleus is higher. Accordingly, the atom remains in an excited state with the energy $E_A$ and two holes in the inflated electron shell \cite{FK}. Accordingly, instead of (\ref{Q}), the effective energy release is realized
      \be
      Q_\text{eff}=M_1-M_2 - E_A = Q - E_A\,.    \label{defect}
      \ee                           
      The process is energetically possible at $Q>0$, but $ Q_\text{eff}$ can also be negative: the excessive energy can be either added to or subtracted from the energy of the satellite quantum. It is $ Q_\text{eff}$ that acts as the resonance defect  $\Delta = |Q_\text{eff}|$.
      We write the formula for the resonance mechanism in the pole approximation using the traditional model. The formula is obtained (cf., for example, \cite{elis}) by multiplying the squared amplitude of the capture itself, $\Gamma_{2e}$, which plays the role of the formation of the doorway state, by the Breit---Wigner resonance factor
\be 
\Gamma_{2e}^{(\gamma)} = \Gamma_{2e} B_W \,,  \label{Wr}
       \ee
      where
\be 
B_W = \frac{\Gamma/2\pi}{\Delta^2+(\Gamma/2)^2}\,.
\label{BW} 
\ee 
In Eq. (\ref{BW}), $\Gamma$ is the width of the
inflated state of the daughter atom with the two holes. In Ref.
\cite{elis} it was taken as the total width of the both holes. This
is not correct: one must also add the width of the final state
\cite{land4}. As a result, the value of $\Gamma$ at least doubles
\cite{FK}. However, the decay of $^{164}$Er was not considered in
Ref. \cite{FK}. For this reason, we make comparison with Ref.
\cite{elis}, keeping their  $\Gamma$ value. The shake-off
contribution is independent of the $\Gamma$ value.

      \begin{figure}[!bt]
      \centerline{ \epsfxsize=10cm\epsfbox{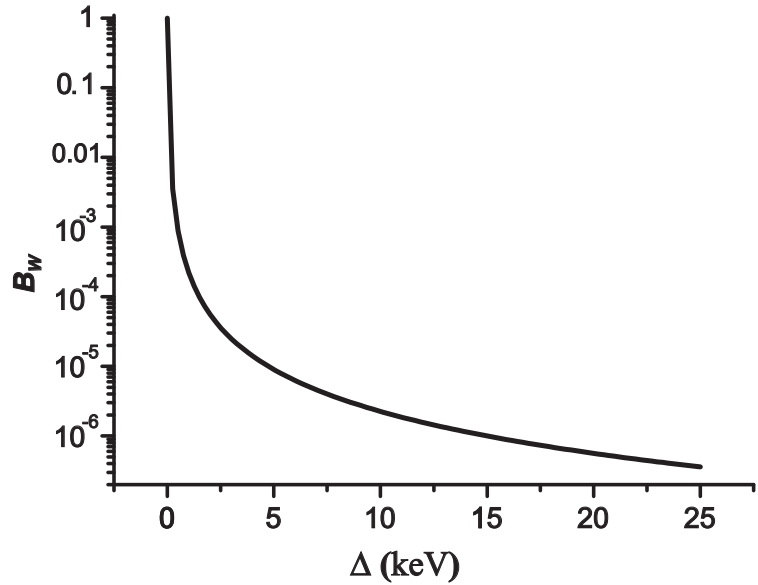}}
      \caption{\footnotesize { Typical dependence of the Breit---Wigner  factor (\ref{BW}) (relativistic units) on the resonance defect $\Delta$.}} \label{Deltaf}
      \end{figure}
\begin{figure}
\centerline{\epsfxsize=7cm\epsfbox{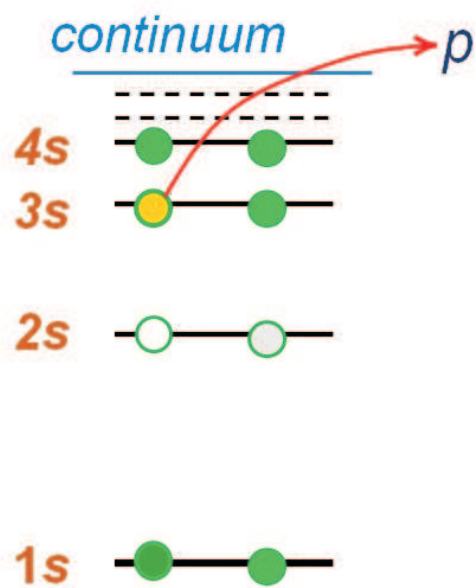}}\caption{\footnotesize Scheme of a representative shake-off process: the $L_1L_1$ electron capture creates two holes in the $2s$ shell. Changed instantaneously inner-atomic electrostatic potential expels the 3$s$ electron into continuum with momentum $p$.}
\label{shakef}
\end{figure}

      A typical value is $\Gamma \approx$  30 eV. For illustration, the scale of the variation of the $B_W$ factor is shown in fig. \ref{Deltaf} {\it versus} the resonance defect. It decreases by a factor of 2 at $\Delta$ = 1.5 keV. Only one candidate is known with such  a value: $^{152}$Gd, which has $\Delta$  = 0.91 keV. In other  cases under discussion, $\Delta$  is at least from a few keV to one or two tens of keV, while the Breit---Wigner factor drops to six orders of magnitude.
      Shake-off is energetically possible only for positive $Q_\text{eff} > 0$, and from the shells whose ionization potential  $I_i$ in the daughter atom (with the two vacancies in the electron shell) satisfies the condition
      \be
      I_i  < Q_\text{eff} \,.                   
      \ee
Then the energy of the shake-off electrons is defined  by their
difference as follows: 
\be E_{sh}= Q_\text{eff}-I_i \,. \ee
      The shake-off arises due to the very rapid, instantaneous change in the inner-atomic electrostatic potential $V_Z(r)$ in the initial atom to the potential in the daughter atom $V_{Z-2}(r)$, see Fig. \ref{shakef}. 
Therefore, the electron wave functions of the initial and final atoms are non-orthogonal, even with the same quantum numbers. To calculate the shake-off probability, most important is the non-orthogonality of the one-electron wave function $\phi_i(r)$ in the parent atom and $\psi_f(r)$ in the daughter atom. Let's denote the change in the potential $\Delta V(r) \equiv V_Z(r)-V_{Z-2}(r)$. Thus, the wave function $\phi_i(r)$  belongs to the parent neutral atom in the ground state, and $\psi_f(r)$ is calculated in the field of the daughter atom with three vacancies: two in the places of the captured electrons and one in the place of the emitted shake-off electron. Then the shake-off amplitude reads as follows \cite{land}:
\be F_{\text{sh}}(\Delta) = \langle \psi_f|\phi _i\rangle  \,.
\label{sh1} \ee 
Similarly to (\ref{Wr}),    one can put down the
full amplitude factorized as  follows:         
\be 
F_{2e}^{\text{(sh)}} = F_{2e}F_{\text{sh}}(\Delta)  \,.    \label{W2bs} \ee
Comparing Eqs. (\ref{W2bs})  to (\ref{Wr}), one arrives at the
following expression for the relative correction to the decay
probability: 
\be
G=\Gamma_{2e}^{(sh)}/\Gamma_{2e}^{(\gamma)}  = \sum_i N_i
|\langle \psi_f|\phi_i\rangle  |^2 / B_W \equiv  \sum_i N_i
|F_{\text{sh}}|^2 /B_W \,,  \label{rtio} 
\ee
 with  $N_j$ being the occupation numbers.

      Within the framework of the resonance fluorescence mechanism, the main contribution comes from the capture of the two $L_1$ electrons in the $^{164}$Er atoms. The capture of lower electrons is energetically forbidden, that of higher ones is suppressed by the decrease in their density at the nucleus together with a decreasing  Breit---Wigner factor, brought about by an increase in $Q_\text{eff}$. As for the nonresonance shake-off mechanism, a decrease in the electron density at the nucleus, for example, in the capture from the $M1$ shell, on the contrary, is partially compensated by an increase in $Q_\text{eff}$, since the shake-off channel from the $L_1$ shell opens. This leads to the fact that, as we shall see, the probability of capture from higher shells gets even higher than the probability of the traditional resonance mechanism. Let $ik$ capture occur from the higher $i$, $k$ shells. Then the acceleration factor can be calculated in comparison with the most probable resonance $L_1L_1$ capture by means of the formula
      \be
      G_{ik} = \frac{\rho_i(0)\rho_k(0)}{\rho_{L_1}^{\ 2}(0)}
      \sum_j N_j |F_{\text{sh}}^{(j)}(|Q_\text{eff}^{(ik)}|)|^2 /B_W \,.  \label{rth}
      \ee

      In Eq. (\ref{rth}), the summation is  carried out over all shells $j$ where  shake-off is energetically allowed. $F_{\text{sh}}^{(j)}(|Q_\text{eff}^{(ik)}|)$ is still the overlap integral of the wave functions of the electron in the initial shell $j$ and the electron in the continuum, but calculated for the actual energy release $Q_\text{eff}^{(ik)}$ corresponding to the $ik$ capture. In the case of the most probable $L_1L_1$ capture, the lowest shell  where the shake-off  occurs is the $M$ shell. Alternatively, if the capture of one of the electrons occurs from the $M$ shell, then the value of $Q_\text{eff}^{(LM)}$ increases by the difference of the ionization potentials of the $L$- and
$M$-shells. This automatically opens the shake-off channel from the
$L$-shell ($L_1$, $L_2$, $L_3$), which leads to a stepwise increase
of the shake probability.

      \section{Results of calculation}

      Calculations by means of Eqs.  (\ref{rtio}),  (\ref{rth}) were performed using the RAINE software package \cite{RAINE,AD}. 
The one-electron wave functions  and their eigenvalues were calculated by means of the self-consistent Dirac---Fock method. In order to better understand the physics of the process, matrix elements (\ref{sh1}) were calculated for a number of hypothetical values of $\Delta$ from 0.05 to 20 keV for all electrons whose ionization potentials are less than the given $\Delta$ value and which, therefore, contribute to the amplitude of the shake-off mechanism of the process. The total widths of the electron hole states are taken from \cite{AD}.

      The calculation results are shown in Figs. 2 -- 5, as well as in tables 1, 2. Our wavefunctions are normalized at unity for discrete states and  the $\delta$ function on the energy scale --- in the continuum. Therefore, the square of the matrix element $F_{sh}(\Delta)$ acquires the dimension reciprocal of the energy. The matrix elements are presented below in the relativistic system of units.
      \begin{figure}[!bt]
      \centerline{ \epsfxsize=10cm\epsfbox{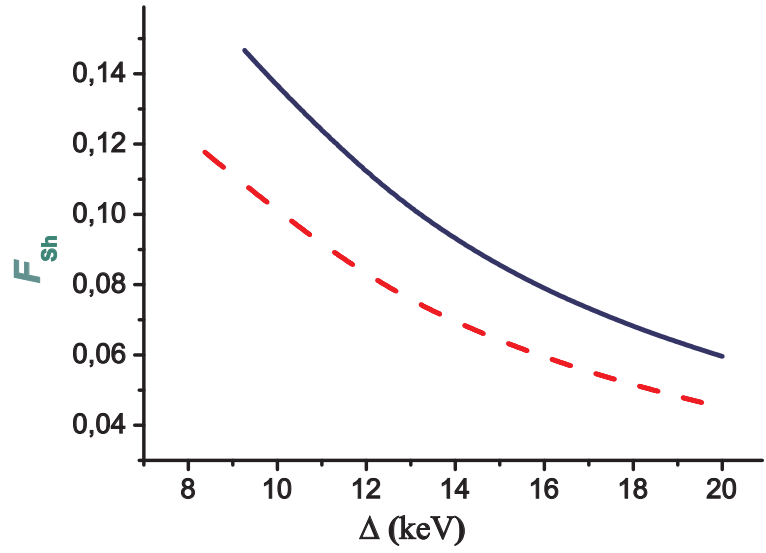}}
      \caption{\footnotesize {Matrix elements $F_{sh}$ for the $2p_{1/2}$- (solid curve) and
      $2p_{3/2}$- (dashed curve) subshells of  $^{164}$Dy atoms against the resonance defect $\Delta$.}} \label{2pf}
      \end{figure}
      The closer the shell is to the core, the greater its contribution to the shake-off, if it is not energetically forbidden. This is illustrated in Fig. 2, which shows the $F_{sh}$ matrix elements for the  $L_2$ and $L_3$ subshells. The curves start from different thresholds: 9.264 and 8.358 keV, respectively. Both thresholds are higher than the effective energy release; therefore, in this case, neither of the curves contributes to shake-off in the most probable case of the $L_1L_1$ capture. Fig. 3 shows matrix elements for the $2p_{1/2}$ -- $5p_{1/2}$ shells. The matrix elements for the rest of the shells are illustrated in fig. \ref{restf}.
      \begin{figure}[!bt]
      \centerline{ \epsfxsize=10cm\epsfbox{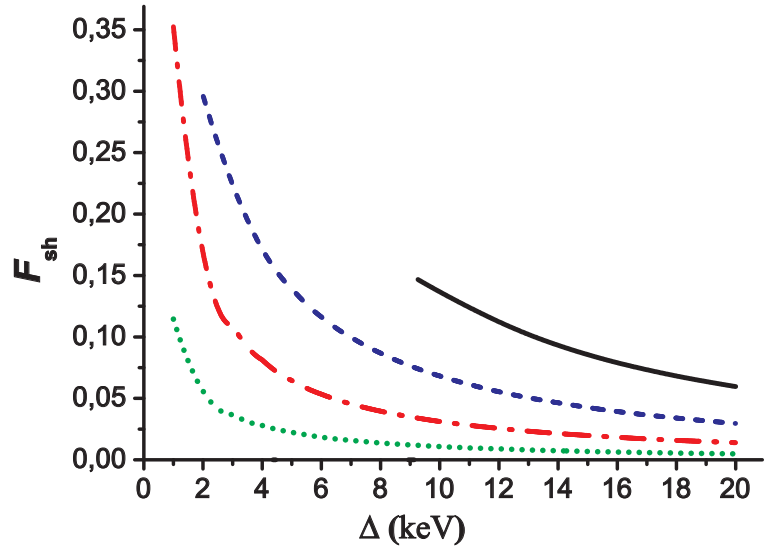}}
      \caption{\footnotesize {Matrix elements $F_{sh}$ for the $np_{1/2}$-subshells of $^{164}$Dy atom: $n$=2 (solid line), $n$=3 (dashed line), $n $=4 (dash-dotted line) and $n$=5 (dotted line).}} \label{npf}
      \end{figure}
      \begin{figure}[!bt]
      \centerline{ \epsfxsize=10cm\epsfbox{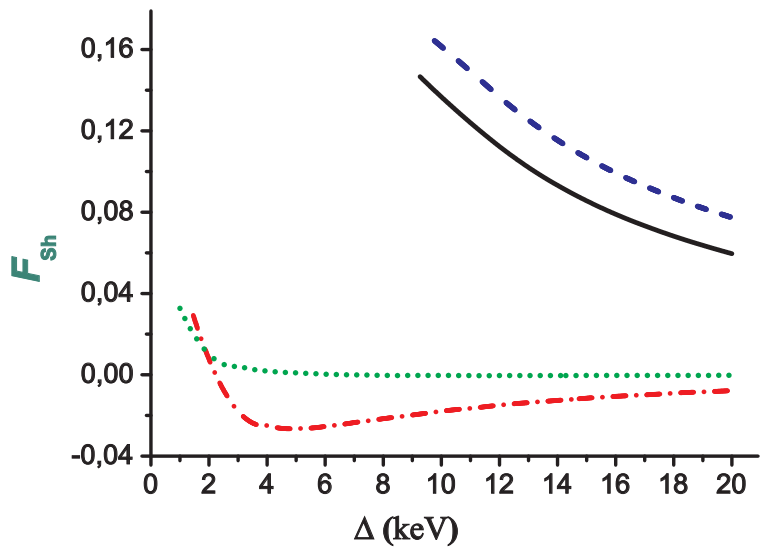}}
      \caption{\footnotesize { Comparison of shell contributions to the shake amplitude as a function of the orbital angular momentum. Matrix elements $F_{sh}$ for the $2s$ subshell (dashed line), $2p_{1/2}$ subshell (solid curve), $3d_{3/2}$- (dash-dotted line), and
      $4f_{5/2}$- (dotted line) subshells of $^{164}$Dy atom.}} \label{restf}
      \end{figure}

\begin{figure}[!bt]
      \centerline{ \epsfxsize=10cm\epsfbox{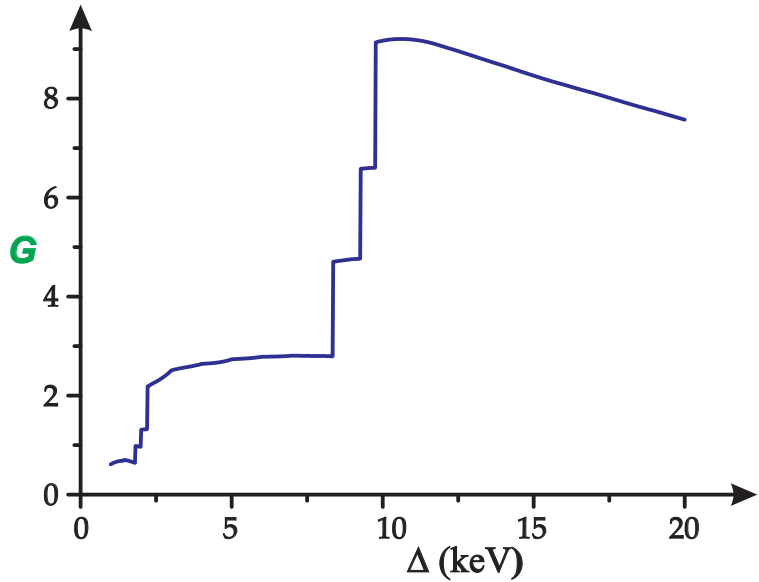}}
      \caption{\footnotesize{The gain of the shake-off mechanism $G$  (\ref{rtio}) as compared to the resonance-fluorescent mechanism in the probability of double neutrinoless $L_1L_1$-capture in $^{164}$Er, depending on the resonance defect  $\Delta$.}}
      \label{gainf}
      \end{figure}

      The total acceleration factor corresponding to the shake-off mechanism from all the electrons, brought about by the $L_1L_1$ capture,  relative to the resonance mechanism is shown in Fig. \ref{gainf}. The probability of this process has a pronounced stepwise character due to the fact that with increasing $Q$, deeper and deeper shells are switched on, and the deeper the shell lies, the greater its contribution at the threshold. As expected, the main contribution comes from the $s$- and $p$-electrons. It can be seen that at small $Q$, the resonance mechanism dominates. At the actual value of $Q$ = 6.82 keV, the contribution of the nonresonance mechanism is three times as  high as  that of  the traditional mechanism.

      Shown in figs. \ref{2pf} -- \ref{restf} values can be used in order to estimate the shake-off contribution in the cases of electron capture from the other, higher shells. Using the total width of the $L_1$ hole in the Dy atom: $\Gamma_{L_1}$ = 4.3 eV \cite{AD}, we obtain by means of formula (\ref{rth})  the total acceleration factors for capture from the $L$, $M$, and $N$ shells. They are presented in Table 1. As one can see from the Table, taking into account higher shells leads to an additional increase of the capture rate by 1.8 times. And the total gain is 5.6 times. Supposing that it is only the shake-off mechanism which makes $e$ capture from higher orbits significant, one can estimate from Table 1 the total fraction of these captures to be of about one third. One can easily calculate the mean shake-off probability: 82\% per capture.
      \begin{table}
      \caption{\footnotesize Partial gains $G_{ik}$ (\ref{rth}) from the shake-off mechanism calculated for $e$ capture in the various shells $ik$. $\rho(0)$ is the normalized at the $L_1L_1$ capture product of the densities \cite{band} of the both captured electrons at the origin }
      \begin{center}
      \begin{tabular}{||c||c|c|c||}
      \hline  \hline
      Shell &   $\Delta$ (keV) &    $\rho(0)$ & $G$ \\
      \hline
      $LL$  &   6.82  & 1 & 2.81 \\
      $LM$   &  14   &  0.218 & 1.22   \\
      $MM$   &      21   &  0.048  &    0.20  \\
      $LN$   &      15.6   &    0.051   &   0.29  \\
      $MN$   &  22.6   &    0.011    &  0.05   \\
      NN  & 24.3  & 0.003  &    0.01  \\
      \hline
      Total:   &    ---  &  ---  &  4.58  \\
      \hline  \hline
      \end{tabular}
      \end{center}
      \label{shtab}
      \end{table}

       \section{Discussion of the results and future prospects}

1. Allowance for shake-off processes in the neutrinoless double
nuclear $e$-capture shifts the theoretical estimates much closer
to the experimental possibilities. Its peculiarity is that
shake-off  renders obsolete the requirement of resonance between the
initial and final atoms. Therefore, it can be expected that it
will be more effective in the cases of nuclei with big
$Q_\text{eff}$ values, when traditional
resonance fluorescent mechanism is suppressed. 
Consideration of the shake-off mechanism significantly refines the half-life estimate. The above calculations confirm this
assumption: taking into account the new mechanism increases the
estimate of the double capture probability by a factor of about 6
in the case of $^{164}$Er. Note that theoretical half-life
relative to the $2e0\nu$ capture scales with $(1/m_\nu)^2$, where
$m_\nu$ is the effective neutrino mass. Taking into account a traditional 
estimate of the half-life of this nucleus 
as $2 \times 10^{30}$ years for $m_\nu$ = 1 eV \cite{elis},
we obtain a refined estimate of the half-life, taking into account the shake-off mechanism, as $T_{1/2}^{0\nu}
\approx 3.6\times10^{29}$  (1 eV/$m_\nu)^2$ years. In
other cases of heavier nuclei with an effective energy release of
$\gtrsim$10 keV, including $^{180}$W, the gain achieves already a
full order of magnitude. 
We summarize the expected results for the half-lives of
the above three candidates: $^{152}$Gd, $^{164}$Er and $^{180}$W in Table
\ref{t3}.
In view of  that the decay period of another candidate for
measuring the $2e0\nu$ capture of $^{152}$Gd remains four orders
of magnitude shorter, we can conclude that it remains a
more likely candidate for setting up an experiment than
$^{164}$Er. 
At the same time, the expected lifetime of $^{180}$W
with respect to the $2e0\nu$ capture turns out to be only 4 times
longer than that of $^{152}$Gd. This can make it a better
candidate, given that the abundance of the $^{152}$Gd isotope in
nature is only 0.2\%.
\begin{table}
\caption{\footnotesize Resulting halflives of $^{152}$Gd, $^{164}$Er
and $^{180}$W double neutrinoless $e$ capture, taking into account both mechanisms} 
\begin{center}
\begin{tabular}{||c||c|c|c||}
\hline   \hline
Nuclei & $^{152}$Gd$\to ^{152}$Sm & $^{164}$Er  $\to$ $^{164}$Dy & $ ^{180}$W  $\to$ $^{180}$Hf  \\

\hline
Decay channel & KL & LL & KK \\
$\Delta$ (keV) & 0.910 & 6.82 & 12.5 \\
Resonance half-lives (years) & 10$^{27}$ & $2\times 10^{30}$ & 
$3\times 10^{28}$ \\
Shake-off half-lives (years) & $8\times 10^{26}$ & $3.6\times 10^{29}$ & 
$3\times 10^{27}$ \\
\hline    \hline
\end{tabular}
\end{center}
\label{t3}
\end{table}

2. Shake-off leads to a radical change of the fluorescence spectrum. 
As mentioned in the Introduction, the appearance of two
satellites at each fluorescence line comprises a characteristic feature of $2e$ capture. First, the energies of the satellites receive an additional shift due to availability of the third vacancy on the place of the shake-off electron. Second, in the case of capture from higher orbits, new satellite fluorescence photons  arise. In the case of $^{164}$Er, in one third of the cases, the capture occurs from higher orbits than $L_1$. Consequently,  satellites  fluorescence quanta appear, which correspond to the transitions of electrons to the states of the $M$- and $N$-shells. This new phenomenon must be taken into account in experiments. One  can
use it for the purpose of  more reliable identification of the
process and its mechanism. A more detailed analysis of the appearing satellite spectrum can be performed elsewhere. 

	3. Allowance for shake-off is also important in investigation of $\beta$ decay and other $\beta$ processes. 
Such studies are carried out with the beta decay of tritium
\cite{triti} in the KATRIN experiment aimed at measuring the mass
of an electron antineutrino and searching for sterile neutrinos.
Shake-off modifies  spectrum of emitted electrons near the upper
bound --- in the region which is most sensitive
to the experimental determination of the neutrino mass.  Restrictions on the mass of the Majorana neutrinos are also
established in the study of double beta decay
\cite{130Te,136Xe,100Mo} and Refs. cited therein. Shake-off
processes are of great importance for correct analysis of the
experimental lifetimes. 

       Similar measurements are also carried out
by studying the $e$-capture at $^{163}$Ho, $^{159}$Dy. Preference is given to these sources because of  the minimum $Q$ value. The lower the $Q$ value,
the greater the number of events falls close to  the upper bound of the spectrum. 
In this case, the neutrino spectrum can be measured by detecting the
secondary processes accompanying  formation of the vacancy:
fluorescence photons, Auger electrons by the calorimetric method.
The calorimetric spectrum of $e$-capture in $^{159}$Dy to the 363.5449 keV level of $^{159}$Tb, $Q$ = 1.14(19) keV was
calculated in Ref. \cite{newCand} with no allowance for shake-off. The calculation was performed
within the Vatai approximation, in which the remaining electrons
of the daughter atom inherit the quantum numbers of the parent
atom. A continuous calorimetric spectrum was supposed to be created by the
width of the formed vacancy. The main contribution near the upper
bound was obtained  due to the $N_1$-capture with the formation of an excited
state corresponding to the configuration [Xe]$4s^{-1}4f^{10}6s^2$.
However,  one can expect a significant contribution from the
shake-off, for example, with the formation of a daughter atom in
a configuration [Xe]$4s^{-1} 5s^{-1}4f^{10}6s^2$. Then the continuity of
the calorimetric spectrum is provided by the escape of the $5s$
electron into the continuum. Moreover, a significant contribution
can also be expected from the electron shake-off from the $N_2$ --
$N_5$ and  higher shells. As a result, the total
contribution of the shake-off can be comparable or  greater
than the contribution from the resonance fluorescence mechanism.

      Summing it up, one can conclude that the nonresonance shake-off mechanism of double neutrinoless $e$-capture is an important example where this process, being of great interest in itself,  manifests itself surprisingly clearly. Search for such surprising manifestations in other $\beta$ processes seems to be a challenging task of the contemporary investigation.

      One of the authors (FFK) would like to thank Yu. N. Novikov for initiating discussions.

\end{document}